\documentclass{elsart}
\usepackage{graphicx}
\usepackage[latin1]{inputenc}

\begin{document}

\begin{frontmatter} 
\title
{Fragment properties of fragmenting heavy nuclei produced in central and
semi-peripheral collisions.}

\author[IPNO,GANIL]{E.~Bonnet} \ead{bonnet@ganil.fr},
\author[IPNO]{B.~Borderie\corauthref{cor}},
\corauth[cor]{Corresponding author - borderie@ipno.in2p3.fr}
\author[IPNO]{N.~Le~Neindre\thanksref{PA}},
\author[IPNO]{M.~F.~Rivet},
\author[LPC]{R.~Bougault},
\author[GANIL]{A.~Chbihi},
\author[CEA]{R.~Dayras},
\author[GANIL]{J.~D.~Frankland},
\author[IPNO,ARTS]{E.~Galichet},
\author[IPNO,LAVAL]{F.~Gagnon-Moisan},
\author[IPNL]{D.~Guinet},
\author[IPNL]{P.~Lautesse},
\author[IFJ]{J.~{\L}ukasik},
\author[LPC,IPNL]{D.~Mercier},
\author[LPC,NIPNE]{M.~P\^arlog},
\author[NAP]{E.~Rosato},
\author[LAVAL]{R.~Roy},
\author[GSI]{C.~Sfienti},        
\author[NAP]{M.~Vigilante},
\author[GANIL]{J.~P.~Wieleczko} and
\author[ASI]{B.~Zwieglinski}
\begin{center}
INDRA and ALADIN Collaborations
\end{center}
\address[IPNO]{Institut de Physique Nucl\'eaire, CNRS/IN2P3, Universit\'e 
Paris-Sud 11, F-91406 Orsay cedex, France.}
\address[GANIL]{GANIL, CEA/DSM-CNRS/IN2P3, B.P.~5027, F-14076 Caen cedex,
France.}
\address[LPC]{LPC, CNRS/IN2P3, ENSICAEN, Universit\'e de Caen, F-14050 Caen 
cedex, France.}
\address[CEA]{IRFU/SPhN, CEA Saclay, F-91191 Gif sur Yvette cedex, France.} 
\address[ARTS]{Conservatoire National des Arts et M\'etiers, F-75141 Paris 
cedex 03, France.}
\address[LAVAL]{Laboratoire de Physique Nucl\'eaire, D\'epartement de Physique, 
de G\'enie Physique et d'Optique,
Universit\'e Laval, Qu\'ebec, Canada G1K 7P4.}
\address[IPNL]{Institut de Physique Nucl\'eaire, CNRS/IN2P3, Universit\'e 
Claude Bernard Lyon~1, F-69622 Villeurbanne cedex, France.}
\address[IFJ]{Institute of Nuclear Physics IFJ-PAN, PL-31342 Krak{\'o}w, 
Poland.}
\address[NAP]{Dipartimento di Scienze Fisiche e Sezione INFN, Universit\`a
di Napoli "Federico~II", I80126 Napoli, Italy.}
\address[GSI]{Gesellschaft f\"ur Schwerionenforschung mbH, D-64291 Darmstadt, 
Germany.}
\address[NIPNE]{National Institute for Physics and Nuclear Engineering,
RO-76900 Bucharest-M\u{a}gurele, Romania.}
\address[ASI]{The Andrzej Soltan Institute for Nuclear Studies, PL-00681,
Warsaw, Poland.}
\thanks[PA]{Present address: LPC Caen, ENSICAEN, Universit\'e de Caen, CNRS/IN2P3, F-14050
Caen Cedex, France} 

\begin{abstract}
Fragment properties of hot fragmenting sources of similar sizes 
produced in central and semi-peripheral collisions are compared in
the excitation energy range 5-10 AMeV. 
For semi-peripheral collisions a method for selecting  compact 
quasi-projectiles sources in velocity space similar to those of 
fused systems (central collisions) is proposed. 
The two major results are related to collective energy.  
The weak radial collective energy observed for quasi-projectile sources
is shown to originate from thermal pressure only.
The larger fragment multiplicity observed for fused systems and their more
symmetric fragmentation are related to the extra radial collective energy 
due to expansion following a compression phase
during central collisions. A first attempt to locate where the different
sources break in the phase diagram is proposed.
\end{abstract}

\end{frontmatter}

\section{Introduction}

Heavy-ion collisions at intermediate energies offer various possibilities to
produce hot nuclei which undergo a break-up into smaller pieces, which is
called multifragmentation. This phenomenon 
is expected to bring information and constraints on
the phase diagram of nuclear matter through measured fragment
properties~\cite{WCI06}. 
In particular by comparing in detail the properties of fragments emitted
by hot nuclei formed in central and semi-peripheral collisions (i.e. with 
different dynamical conditions for their formation) one can expect
to reveal features which characterize where those hot nuclei break 
in the phase diagram~\cite{Kun95,Natow02,MDA03,Ma05}. It is the final goal
of this article; the first question being: do we see
any different features in fragment properties ?
For reactions at small impact parameters in the Fermi energy domain,
one can select the collisions where the two nuclei merge into a quasi-fused
system (QF) 
after full stopping~\cite{I46-Bor02}.
At larger impact parameters and for higher incident energies,
only a fraction of each nucleus interacts producing
in the outgoing channel quasi-target (QT) and quasi-projectile (QP) sources.
For both cases large energy dissipations occur but some constraints 
applied to nuclei
are different with a compression-expansion cycle for central collisions and
a friction-abrasion process which can produce dynamical emissions in the
contact region (mid rapidity emissions) for peripheral
collisions~\cite{Mon94,Dem96,I17-Pla99,I45-Luk03}. Therefore,
to make a meaningful comparison of fragment properties which can be 
related to the phase  diagram, hot nuclei showing to a
certain extent statistical emission features must be selected.
It is done for central collisions by selecting compact events in velocity
space (flow angle selection). For peripheral collisions a selection method is
proposed and applied to quasi-projectiles.
Hot nuclei with A around 150-200 were produced at GANIL 
in $^{129}Xe$+$^{nat}Sn$
central collisions at five bombarding energies in the range 25-50 A MeV and
at GSI in semi-peripheral Au+Au collisions at 80 A MeV incident energy.

The paper is organized as follows. In section 2 we briefly present the
experimental set-up and conditions. We recall the criteria
allowing to select experimental events corresponding to fused systems.
Then the method employed to extract the excited QPs is described.
Global properties of the different selected sources are finally discussed
using the excitation energy as control parameter. A comparison of
fragment charge partitions associated with the different sources is shown 
in section 3. Section 4 presents a detailed study
of kinetic fragment properties produced in central and peripheral collisions. 
Radial collective energies are compared by means of
fragment relative velocities.
Section 5 is devoted to discussion and conclusions.

\section{Experimental selection}

\subsection{Experimental procedure}
Beams of $^{129}$Xe, accelerated at five incident energies: 25, 32, 39, 45
and 50~A~MeV by the GANIL facility, bombarded  a thin target of natural tin 
(350 $\mu$g/cm$^{2}$). Hot nuclei with A around 150-200 were produced in
central collisions. Nuclei of similar masses  were obtained at GSI in 
semi-peripheral Au+Au collisions at 80 A MeV incident energy.
This energy appears to be a good compromise between high 
energy detection limit for H isotopes introduced by the detectors, 
a grazing angle within the measured angular range 
and a good characterization for QPs which are well separated from QTs in
velocity space. For this experiment the $^{197}$Au beam was impinging on a 
2 mg/cm$^{2}$ $^{197}$Au thick target.  

The data were collected with the 4$\pi$ multidetector INDRA which is described
in detail in~\cite{I3-Pou95,I5-Pou96}. 
INDRA consists of 336 telescopes covering about 90\% of
the 4$\pi$ solid angle.
The configuration used at GSI differed in the composition of the first ring
(2$^{\circ}$- 3$^{\circ}$): 
the phoswich scintillators were replaced by 12 telescopes, each composed of a
300 $\mu$m Si detector followed by a 14 cm long CsI(Tl) scintillator. 
Accurate fragment identification and energy calibration were achieved with
INDRA; the energy of the detected
products is obtained with an accuracy of 4\%. Further details can be found
in~\cite{I14-Tab99,I34-Par02,I33-Par02,I44-Trz03}.

\subsection{Event selection for quasi-fused systems}
\begin{figure}[!hbt]
\includegraphics*[scale=0.70]{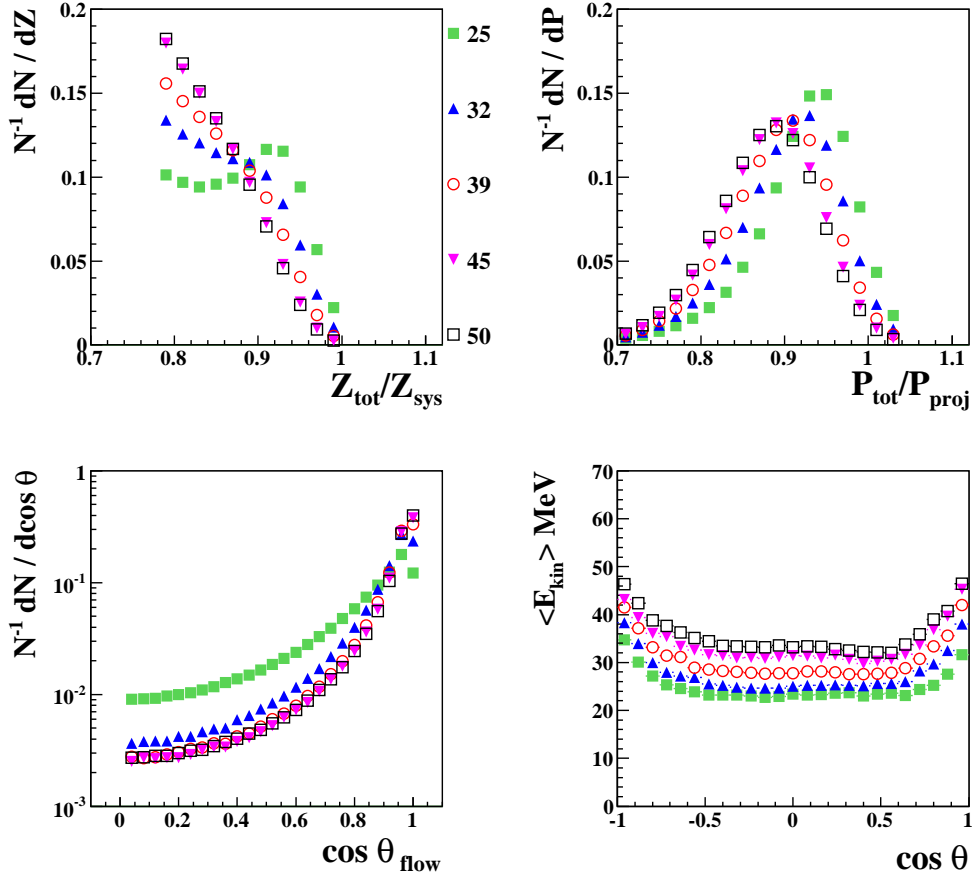}
\caption{Results for Xe+Sn central collisions at 5 energies:
25, 32, 39, 45 and 50 AMeV. Top -  Normalized detected charge (left) 
and pseudo linear momentum (right) distributions of complete events.
Bottom left panel: normalized angular distributions of the CM flow angle, 
for complete events; bottom right panel: mean kinetic energy vs c.m. angle 
for light charged particles ($Z\le4$) for the compact events
($cos \theta_{flow}\le0.5$).}\label{fig1}
\end{figure}
A two step procedure was used to select QF sources. First of all 
``complete experimental events'' were selected by requiring that at least 
80\% of the total charge of the system was measured; the corresponding event 
distributions are displayed in the left upper part of fig.~\ref{fig1}. 
That minimum percentage of the total charge induces a lower limit on the 
total pseudo linear momentum  (P$_{tot}$/P$_{proj}$ - see eq. \ref{ptot}) 
around 70\% of the entrance channel 
value (see right upper part of fig.~\ref{fig1}) and selects central 
collisions only.
\begin{eqnarray}
\label{ptot} P_{tot} = |\sum_{i=1}^{M_{tot}}\vec{\beta_{i}}\gamma_{i}Z_{i}| 
\end{eqnarray}
$M_{tot}$ is the total measured charged product multiplicity in the event.
Then, compact single sources in velocity space were selected by imposing 
the constraint of flow angle 
$\geq 60^{\circ}$~\cite{Lec94,I28-Fra01,I29-Fra01}.
To calculate $\theta_{flow}$ the kinetic energy tensor was built with 
fragments (Z$\geq$5) and starting from fragment multiplicity M$_{frag}\geq1$.
Indeed it was shown in previous studies that while events present the
topology of emission from two sources at small flow angles, they evolve
towards a single-source configuration above $60^{\circ}$ (see figure~9 in 
ref~\cite{I28-Fra01} and figure~1 in ref~\cite{I46-Bor02}). The rather flat
$\cos{\theta_{flow}}$ distribution (variation of about 35\% between
$\theta_{flow}$=60$^{\circ}$ and 90$^{\circ}$) observed for each 
incident energy in the selected range 
(see left bottom part of fig.~\ref{fig1}) indicates a
fragment emission which can be associated with a strong degree of
equilibration. Note that such QF sources are produced at high excitation
energy (above 3.0-3.5 AMeV) and their deexcitation through fission is
suppressed~\cite{MDA03}.

The charged particles (Z = 1-4) to be associated with
the single sources 
were determined from angular and energetic criteria. 
Figure~\ref{fig1} (right bottom part) shows the centre of
mass average energy of particles in coincidence with the single sources
as a function of their centre of mass emission angles.
Large energies are measured forward-backward indicating
preequilibrium emissions whereas rather constant average
values are observed over a 2$\pi$ solid angle (angular
range 60$^{\circ}$-120$^{\circ}$). 
So, assuming isotropic emission, twice the charged particles on
that angular range were associated with the sources event by
event~\cite{I29-Fra01}.

Finally the calorimetric method~\cite{Viol06} was used to evaluate the source 
excitation energy. The following hypotheses have been made: a level density
parameter equal to A/10, the average kinetic energy of neutrons equal to their
emitting source temperature and the Evaporation Attractor Line formula 
(A=Z(2.072+2.32$\times10^{-3}$Z))~\cite{Cha98} applied for fragment
mass determinations. EAL is especially well adapted when heavy fragments
($Z>20$)
result from the deexcitation of the neutron deficient sources produced. More
details on calorimetry are presented in the appendix.
For all incident energies Gaussian distributions are obtained for the 
excitation energies of the source.

\subsection{Selection of quasi-projectile sources}

For peripheral collisions at 80 A MeV incident energy, 
the relativistic fragment (Z $\geq$ 5)
kinetic energy tensor (events with at least one fragment, $M_{frag}\geq1$ - 
eq. \ref{tenseur}) 
was computed in the centre of mass of the reaction to define the ellipsoid
frame of each event, comprising $M^{tot}_{frag}$ fragments. 
\begin{eqnarray}
\label{tenseur} T^{uv} = \sum_{i=1}^{M^{tot}_{frag}}\left(\sum_{u,v=1,3}\frac{P^{(u)}_{i}P^{(v)}_{i}}{
(1+\gamma_{i})m_{i}} \right)
\end{eqnarray}
 QP sources were selected through a completeness criterion 
 on the total detected charge (eq. \ref{zfwd}) and pseudo-momentum 
 (eq. \ref{pfwd}) on the forward part 
of each event - defined as all reaction products 
with a positive rapidity, $M_{fwd}$, in the ellipsoid frame.
\begin{eqnarray}
\label{zfwd} Z_{fwd} = \sum_{i=1}^{M_{fwd}}Z_{i} \qquad Z_{fwd}/Z_{proj}\in[0.80,1.10] \\
\label{pfwd} P_{fwd} = |\sum_{i=1}^{M_{fwd}}\vec{\beta_{i}}\gamma_{i}Z_{i}|
\qquad P_{fwd}/P_{proj}\in[0.60,1.10]
\end{eqnarray}
$Z_{proj}$ and $P_{proj}$ refer to projectile charge and momentum in the
laboratory.

These preliminary selections lead to an ensemble of sub-events
called complete QP events.
Then the velocity of each reconstructed  QP source is determined 
using fragments only.

\begin{figure}[!hbt]
\includegraphics*[scale=0.65]{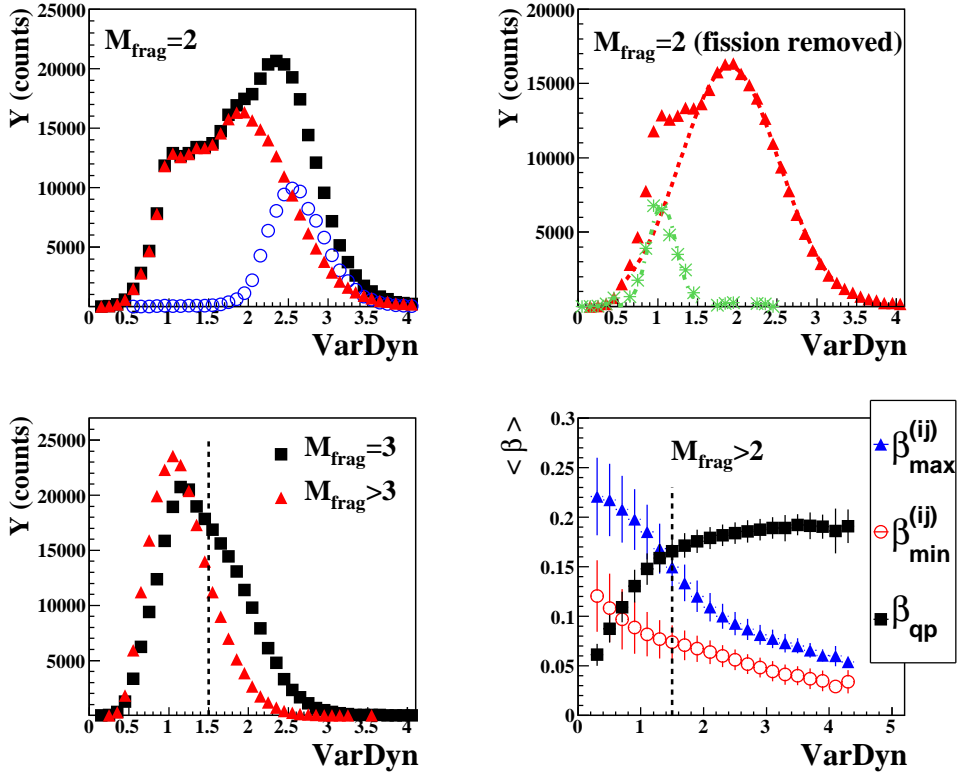}
\caption{Top - Left panel: for QP complete events with $M_{frag}=2$, 
distribution of VarDyn (eq.~\ref{eq_vardyn} - full squares) decomposed 
into fission events ($Z_{1}\times Z_{2}\geq900$ - open circles) and the 
complementary part (triangles). Right panel: same, with the further 
decomposition into compact events (dotted line) and non-compact events 
(stars). Bottom - Left panel:
distribution of VarDyn for events with 3 and more than 3 fragments.
Right panel: for $M_{frag}>2$ QP complete events, mean evolution with VarDyn 
of the minimum and maximum relative velocity between pairs of fragments in
an event, and reconstructed QP velocity; error bars indicate the standard 
deviations of two-dimensional distributions.}\label{fig_vardyn}
\end{figure}
To perform a meaningful comparison between the fragment properties of
single sources defined in the previous section and  QP sources, 
we must minimize the contribution of dynamical emissions for fragments. 
Indeed, in this energy regime and for peripheral and mid-peripheral 
collisions it is well known that a large part of the
collisions end-up in two remnants coming from projectile and target
and some particles and fragments with velocities intermediate between those
of the remnants.
They are called mid-rapidity products. They may have
several origins: direct preequilibrium emission from the overlap region
between the incident partners or a neck of matter between them which may
finally separate from QP or QT, or from both, as well as emissions from QP/QT
partially deformed and locally highly
excited~\cite{I17-Pla99,I45-Luk03,I23-Lef00,I36-Col03,Pia06}.
On the other hand, with higher dissipation associated with the decrease of the 
impact parameter, the velocities of the outgoing partners (QPs and QTs)
are much closer to the centre of mass velocity, which makes difficult
the definition of QP and QT sources.
Therefore a compactness criterion in velocity space was defined and applied
to select compact QP events among complete QP events which comprise at least
two fragments.
With this aim, we investigate further the global energetic properties of 
fragments. We define the VarDyn  observable as follows:
\begin{eqnarray}
\mathrm{VarDyn} = \frac{\beta_{QP}}{\beta_{rel}}\;\mathrm{with}\;\beta_{QP}=|\sum \vec{p^{(i)}}|/\sum
E^{(i)} \nonumber \\
\label{eq_vardyn}
\mathrm{and}\;\beta_{rel}=\frac{2}{M_{frag}(M_{frag}-1)}\sum_{i < j}|
\vec{\beta^{(ij)}}| \qquad \vec{\beta^{(ij)}}=\vec{\beta^{(i)}}-\vec{\beta^{(j)}}.
\end{eqnarray}
$p^{(i)}$, $\beta^{(i)}$ and $E^{(i)}$, which represents the total energy of
fragment $(i)$, are defined in the reaction centre of mass. $M_{frag}$ is
the number of fragments among the $M_{fwd}$ products.
$\beta_{QP}$ is related to the dissipated energy whereas $\beta_{rel}$ gives
a hint of the dispersion of fragments in velocity space. Compact QP sources
should have small values of $\beta_{rel}$ and thus large values of VarDyn.

With this ratio we compare the average
position and distance between fragments with the
reconstructed position of the QP in the velocity space to evidence compact
configurations corresponding to events with fragments localized 
around the projectile velocity (larger values of VarDyn). 
Let us focus first on events with $M_{frag}$= 2 ( 27\% of complete events).
The distribution of VarDyn is plotted in the left upper part of figure
\ref{fig_vardyn}: it exhibits a peak for VarDyn=2.4 and a pronounced shoulder 
for values around 1-1.5. It has long been known that excited Au nuclei will
undergo symmetric fission. Fission of Au QPs was characterized
 by using the criterion $Z_{1}\times Z_{2}\geq900$~\cite{I61-Pic06} (6.3\%
of complete events). The corresponding distribution of VarDyn (open circles)
has a Gaussian shape centered around VarDyn=2.5. Removing the
fission contribution, the complementary part still presents a Gaussian shape
centered around VarDyn=1.8 with a shoulder for VarDyn$<$1.5.
The right upper part of figure~\ref{fig_vardyn} shows the final 
decomposition with two components (after removing fission): compact QP events 
(black triangles) and QP events with mid-rapidity fragments (stars), 
associated with small values of VarDyn. That figure shows that
VarDyn$>$1.5 is a good  criterion to select compact QP events with two
fragments. 

For events with $M_{frag}>2$, the direct dependence between impact parameter, 
dissipation, fragment production and velocity of the QPs makes the separation 
between the two classes of events impossible just starting from the observed 
VarDyn distributions (see left lower part of figure~\ref{fig_vardyn}).
To test the lower limit previously deduced for $M_{frag}$=2, we 
introduce the minimum, $\beta^{(ij)}_{min}$, and maximum,
$\beta^{(ij)}_{max}$,
relative velocity between pairs of fragments
calculated for each event. 
The right lower plot of figure~\ref{fig_vardyn} shows the evolution of their
mean values $<\beta^{(ij)}_{min}>$ and $<\beta^{(ij)}_{max}>$ and the
evolution of $<\beta_{QP}>$ with VarDyn.
$<\beta_{QP}>$ increases rapidly up to VarDyn around 1.5 and then evolves
much more gently, as it is limited by the projectile velocity
$\beta_{proj}=0.2$.  For VarDyn values above 1.5,
$<\beta^{(ij)}_{max}>$ becomes lower than $<\beta_{QP}>$, which indicates 
 source velocities sufficiently different from the reaction centre of mass 
velocity. 
On the other hand the difference
between $<\beta^{(ij)}_{min}>$ and $<\beta^{(ij)}_{max}>$ remains small
above VarDyn around 1.5 whereas it increases for lower VarDyn values. 
That ensemble of observations confirms that VarDyn=1.5 is a good minimum 
condition to select compact QP events irrespective of the 
fragment multiplicity.
\begin{figure}[!hbt]
\includegraphics*[scale=0.65]{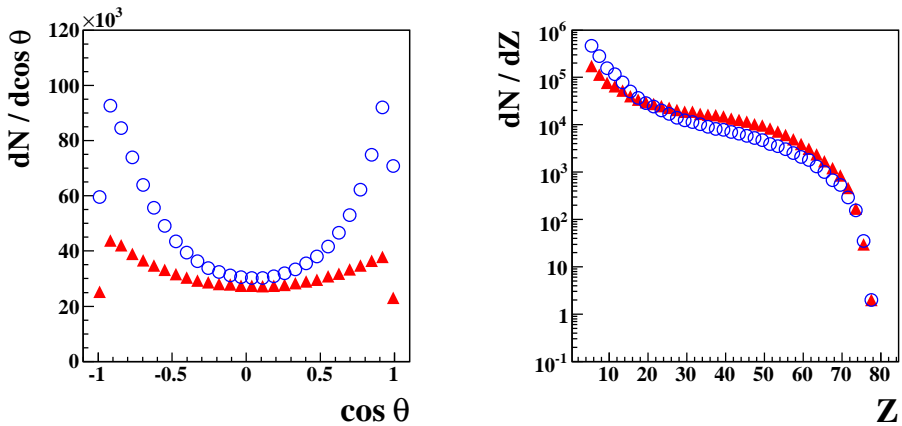}
\caption{\it Fragment angular (left panel) and charge
(right panel) distributions
for compact (triangles) and rejected
non-compact (open circles) events with $M_{frag}\ge2$ and fission events
removed.}\label{vpvp_final}
\end{figure}
Let us come now to a comparison between selected and rejected events. 
Figure~\ref{vpvp_final} shows, for $M_{frag}\ge2$
and $Z_{1}\times Z_{2}<900$ events, the fragment angular
distributions in the QP frame. One notes the much flatter distribution 
associated with compact events, which indicates a fragment emission closer to 
isotropy. Information on fragment size distributions is also displayed in 
the right part of the figure. The rejected events comprise  more 
 light fragments ($Z<20$) than the compact ones.

To summarize, the second and third rows of Table~\ref{stat} indicate 
the proportion of the different events at this stage. Compact events, 
without fission events, correspond to 61.2\% of complete events.
 

\begin{table}[!hbt]
\begin{center}
\begin{tabular}{|c|c|c|c|c|}
\hline
& $M_{frag}=1$
& $M_{frag}\ge2$
& Fission
& Total
\\\hline
Complete
& 585885 (39.7\%)
& 797595 (54.0\%)
& 93339 (6.3\%)
& 1476819 (100\%)
\\\hline
Compact
& 585885 (58.8\%)
& 317986 (32.0\%)
& 91840 (9.2\%)
& 995711 (67.4\%)
\\\hline
Fission removed
& 585885 (64.8\%)
& 317986 (35.2\%)
& - 
& 903871 (61.2\%)
\\\hline
Size selection
& 383730 (81.9\%)
& 84580 (18.1\%)
& - 
& 468310 (31.7\%)
\\\hline
\end{tabular}
\caption{Summary of the percentage of different QP events. The second, third 
and fourth columns refer to events with fragment multiplicities 
 $M_{frag}=1$, $M_{frag}\ge 2$ and $Z_{1}\times Z_{2}<900$ 
 and to fission events ($Z_{1}\times Z_{2}>=900$). 
At each selection step the last column gives the percentages of kept events
relative to the number of complete events.}
\label{stat}
\end{center}
\end{table}

\begin{figure}[!hbt]
\includegraphics*[scale=0.65]{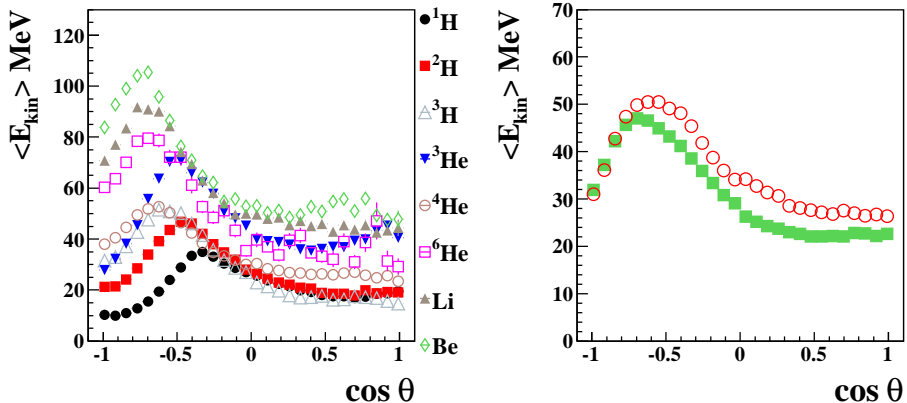}
\caption{ Left: average kinetic energy vs emission angle in the QP frame of
the different light charged particles associated with compact QP sources 
with $<E^*>$=6.3 AMeV. Right: same but averaged over all charged particles
for QP sources with $<E^*>$=4.8 (lower curve) and 7.8 AMeV 
(upper curve).}\label{fig04}
\end{figure}

The light charged particles ($Z<5$) with positive rapidity in the ellipsoid
frame have different origins:
mid-rapidity dynamical emissions, pre-equilibrium emissions due to the 
limited overlap between projectile and target and statistical emissions 
from QPs and QTs. The evolution of their mean kinetic energy as a function 
of their emission angle in the QP frame 
is illustrated in fig.~\ref{fig04} (left panel).
The right panel of the figure shows, averaged over all charged particles,
the same trend for different excitation energies of QP sources (for the
calorimetry procedure - see just after).
They clearly show a flatter behaviour for forward angles
in the QP source frame. To estimate the contribution of the QP statistical
emission we have adopted the  method  already
used in~\cite{I61-Pic06} which consists in keeping only the particles emitted 
forward in the QP frame, and doubling their contribution (charge, mass and 
energy) assuming a forward-backward symmetric emission.

Note that a large part of the emitted particles (a factor of 3 in 
multiplicity) are localized in the backward part of the sources. 
As a consequence, starting from a QP detected charge $Z_{fwd}$ 
constrained by the completeness criterion, this asymmetry 
backward/forward leads to large distributions for the sizes of the 
reconstructed QP sources. To overcome this drawback we have finally 
selected QP source sizes, $Z_s$, which correspond to $Z_{s}$/$Z_{fwd}$ ratios
comprised between 0.9 and 1.0; it is worth noting that the total (rejected)
mid-rapidity charge is on average equal to four for the selected events.
This last size selection withholds 31.7\% of
complete QP events (bottom row of table~\ref{stat}).

The following step consists in the evaluation of the excitation energy of the 
sources. An event by event calorimetry follows the procedure used above for 
QF sources. The hypotheses are identical except for charged particle 
contribution (doubling the forward part) and for fragment masses for which 
we use the formula (A=Z(2.045+3.57$\times10^{-3}$Z))~\cite{Cha98}, 
better adapted for excited nuclei close to the beta stability valley. 
Note that, compared to the EAL
formula, differences for masses appear only for Z greater than 40.
The derived excitation energy range for sources was divided in bins of 0.5
AMeV width for comparisons with QF sources.

In the following, we will only use the excitation energy per nucleon as 
sorting parameter to compare fragment properties from both types of sources.
Whereas the relative values of the excitation energies for each type of source 
should be reliable, one can wonder about the comparison between the
excitation energy scales between QF and QP sources. 
In the next subsection we will present arguments indicating that the two
scales are in agreement within 10\%.

\subsection{Global properties of selected QF and QP sources}
\begin{figure}[!hbt]
\includegraphics*[scale=0.65]{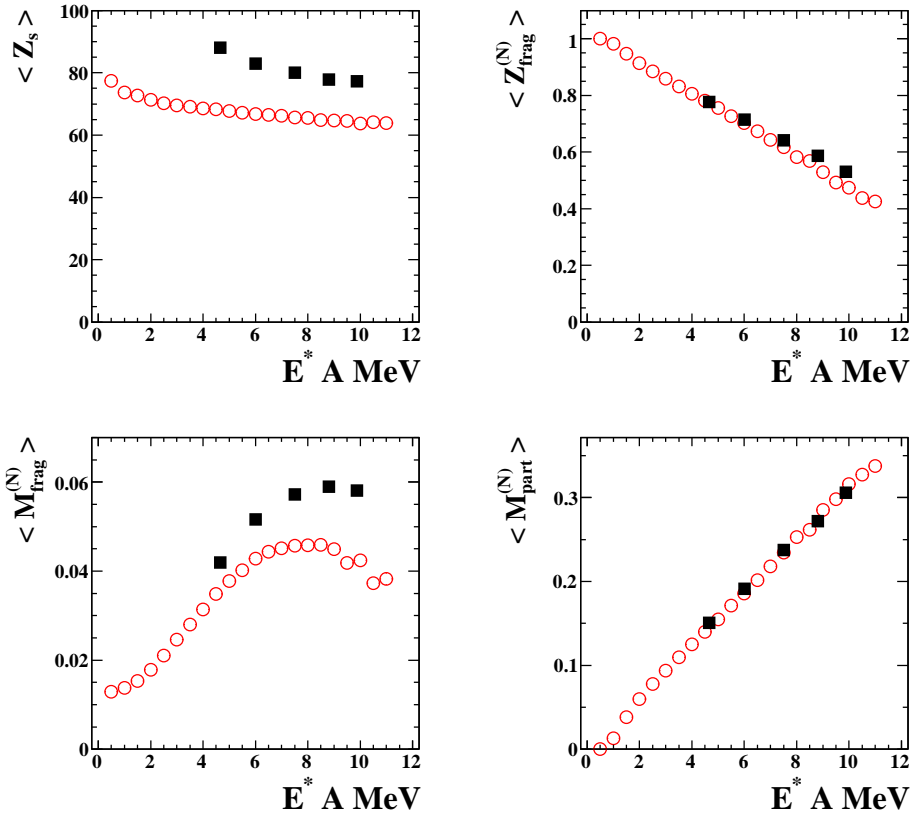}
\caption{Full squares and open
circles stand respectively for QF and QP sources. 
Top: average charge of the sources (left) and of the charge bound in
fragments normalized to the source charge (right) vs the excitation energy
per nucleon. Bottom: average values of the 
normalized fragment (left) and particle (right) multiplicities.}\label{zs_zf}
\end{figure}

The global properties of selected sources are displayed in figure~\ref{zs_zf}. 
In the left upper part of the figure the evolutions of the source 
sizes are plotted as a function of their excitation energy per nucleon. 
The common excitation energy range for the two types of sources is 
[5,10]~AMeV and their variations in charge on that excitation energy 
range are around ten units of charge; the
ratio between the two types of sources is about 1.2-1.3, very close to the
Z ratio between Xe+Sn (104) and Au (79).
The average proportion of charges bound in fragments as a function of the 
excitation energy per nucleon is shown in the upper right hand
side of figure~\ref{zs_zf}. The quantitative evolution of that observable,
normalized to the sizes of the sources, $<Z_{frag}^{(N)}>$, 
is the same for both QF and QP: a linear decrease of charge bound in 
fragments when the excitation energy per 
nucleon increases. This behaviour shows that, for a given excitation energy, 
the sharing among particles and fragments is the same for central
and peripheral collisions, which confirms that multifragmentation is mainly
driven by the energy deposited into the sources~\cite{Tam06}. 
Thus, the knowledge, for a source, of the proportion of charges bound 
in fragments (or the complementary knowledge of the total Z found 
in particles)  provides a good estimate of its excitation energy. 
Note that the proportion varies from about 0.8 to 0.5 in the common 
excitation energy range. Mean fragment multiplicities normalized to the 
size of the source are shown in the lower 
left part of figure~\ref{zs_zf}.
Much higher multiplicities are obtained for QF sources  
than for QP ones but they both present a maximum in the same
excitation energy range (8-10 AMeV). On the other hand the variances of the
normalized multiplicity distributions, which are
not presented here, are very
similar. Finally in the right bottom panel of the figure are displayed the 
normalized multiplicities of particles
$<M_{part}^{(N)}>$, which appear completely similar. 

\section{Fragment charge partitions}
\begin{figure}[!hbt]
\includegraphics*[scale=0.65]{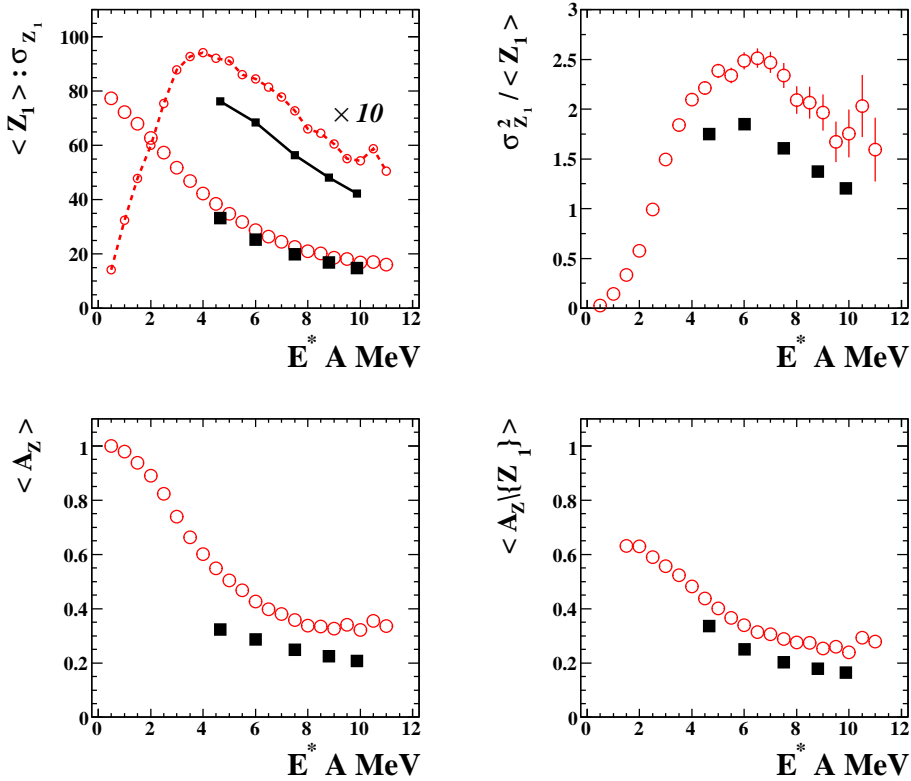}
\caption{ Full squares and open
circles stand respectively for QF and QP sources. 
Top left:  average values and standard deviations 
(multiplied by 10 - smaller symbols joined by lines) of the charge of
the biggest fragment vs the excitation energy per nucleon. Top right:
normalized variances of the charge of the biggest fragment vs the
excitation energy per nucleon. Bottom: evolution
of the charge asymmetry - with (left) and without (right) the biggest
fragment - as a function of the excitation energy per nucleon.}\label{z1}
\end{figure}
As shown previously the percentage  of charge bound in fragments
is the same for QF and QP sources with the same excitation energy.  
How is this bound charge shared among fragments?
A first answer comes from  the properties
of the charge of the biggest fragment ($Z_{1}$), because several studies
point out its specific behaviour~\cite{Des98,I57-Tab05,I61-Pic06} and 
because it is a good candidate for the order parameter of a phase 
transition in hot nuclei~\cite{Bon00,Car02,I54-Fra05,Gul05,Cha07}.
In figure~\ref{z1} (upper left part), the evolutions, with the excitation 
energy, of its mean value and of the associated fluctuations are plotted. 
The mean value appears as mainly governed by excitation energy and is 
largely independent of system sizes and of production modes. This 
effect was already observed in~\cite{I12-Riv98,T41Bon06} for two QF sources
with charges in the ratio 1.5; its occurrence when comparing QF and QP
sources would indicate that their excitation energy scales do agree, within
10\%.
The fluctuations, on the contrary, exhibit sizeable differences. 
In the common energy range, the standard deviations of $Z_{1}$ decrease when 
the excitation
energy increases but they are larger for QP sources. In this latter case
they show a maximum value around 4.5 AMeV which is in good agreement with 
systematics reported for QP sources in~\cite{Gul06,I63-NLN07} and seems
to correspond to the centre of the coexitence region of phase transition~\cite{T41Bon06,Bon08}.
Normalized variances $\sigma^2_{Z_1}/<Z_1>$ are also reported
in figure~\ref{z1} (upper right part); it was
proposed in~\cite{Dor99} that the maximum of that observable indicates the
critical region of the phase transition. We note that, as compared to
standard deviations, normalized variances exhibit a maximum for both QP and
QF sources in the excitation energy region 6-7 AMeV, which seems to correspond to the
gas-like border of the transition and consequently possibly to the critical
region~\cite{I59-Bor04,Bon08}.
A surprising result coming from that comparison between QP and QF 
sources is the difference between the behaviours of the charge bound 
in fragments, $Z_{frag}$ -
fig~\ref{zs_zf} and of $Z_{1}$  at a given excitation energy. 
In the first case one observes a scaling with the size of the
sources whereas the second  exhibits an independent mean value.
How is $Z_{frag}$ partitioned into fragments?
An overview of all information related to fragment charge partition
can be obtained with a new generalized  charge  asymmetry variable
calculated event by event. 
For two fragment events ($M_{frag}$=2), the usual proposed observable
is $a_{12}=\frac{Z_{1}-Z_{2}}{Z_{1}+Z_{2}}$~\cite{Kre93}. To take into account
distributions of fragment multiplicities which differ for the two sources,  
the generalized asymmetry ($A_{Z}$) reads: 

\begin{eqnarray}
\label{eq_az} A_{Z}= \frac{1}{\sqrt{M_{frag}-1}}\frac{\sigma_{Z}}{<Z>}
\end{eqnarray}

This observable evolves from 1 for asymmetric partitions to 0 for equal size
fragment partitions (symmetric). 
For the one fragment events, mainly present for QP
sources, we compute the $A_{Z}$ observable by taking the first particle 
in size hierarchy included in calorimetry. In the left bottom part of 
figure~\ref{z1}, 
the mean evolution with excitation energy of the generalized asymmetry
is shown. Differences are observed which well illustrate how different 
are the repartitions of $Z_{frag}$ between fragments for QF and QP 
multifragmenting sources. QP partitions are more asymmetric in the entire 
common excitation energy range. To be sure that this observation does 
not simply reflect the peculiar behaviour of the 
biggest fragment, the generalized asymmetry is re-calculated for partitions
$M_{frag}>1$, and noted 
$A_{Z} \backslash \{Z_{1}\}$, by removing $Z_{1}$ from partitions
(bottom right panel of figure~\ref{z1}). The difference between the 
asymmetry values for the two source types persists.
A possible explanation of those experimental 
results can be found by looking at kinematic properties of
fragments~\cite{Kun95}.

\section{Mean fragment relative velocities and radial collective energy.}
Radial expansion energy following a compression phase is 
predicted to be present in semi-classical simulations of central collisions 
in the Fermi energy domain~\cite{Mol88,Ber88,Surau89,Bon94}. In experiments
it was obtained, in most of the cases, from
comparisons of kinetic properties of fragments with statistical models. For
QF sources the centre of mass of the reaction is used as the reference frame
to derive kinetic properties of fragments. For QP sources produced in
peripheral and semi-peripheral collisions, the definition of
the QP frame is correlated to the fragment kinetic energies.
Conversely the mean relative velocity between fragments ($\beta_{rel}$
- eq.\ref{eq_vardyn} for $M_{frag}>$1), is independent of the reference 
frame, and can provide information about possible radial collective energy. 

\begin{figure}[!hbt]
\includegraphics*[scale=0.65]{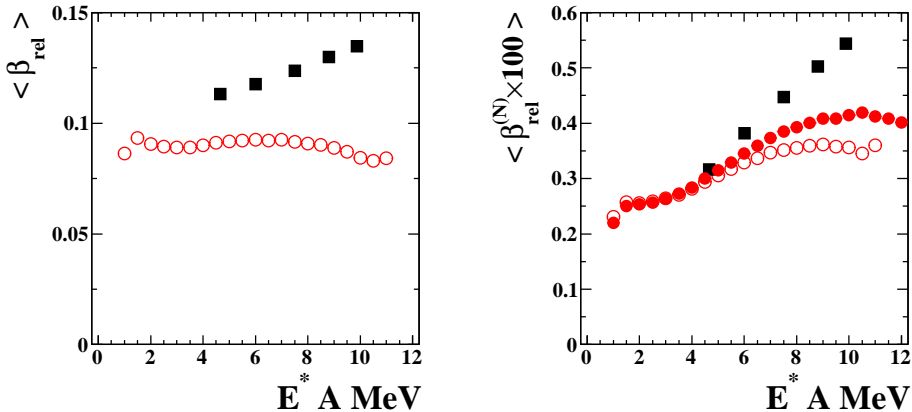}
\caption{ Full squares and open
circles stand respectively for QF and
QP sources. Full circles correspond to QP sources produced in 100 AMeV
collisions. Evolution of the mean relative velocity of fragments,
$\beta_{rel}$, (left panel) and of that normalized observable, 
$\beta^{(N)}_{rel}$, (right panel) with the excitation energy per nucleon. 
}\label{vr}
\end{figure}

\begin{figure}[!hbt]
\includegraphics*[scale=0.55]{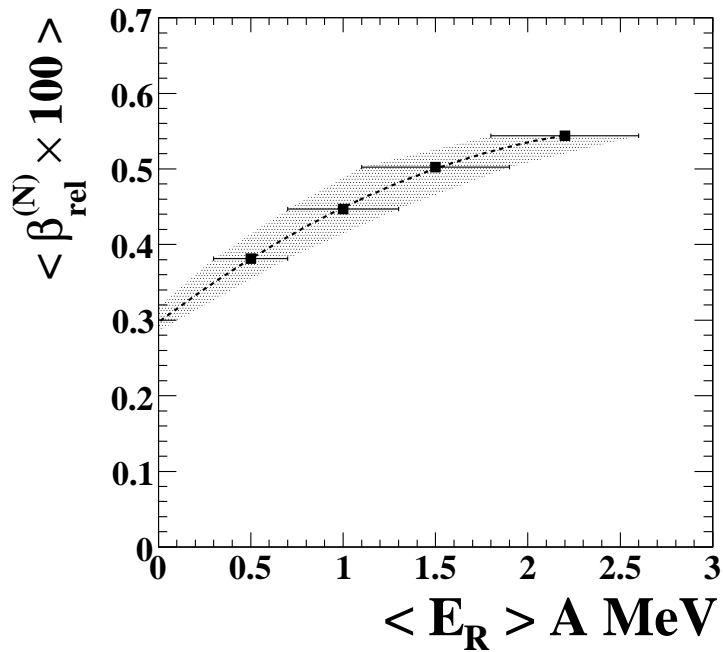}
\caption{Radial collective energy calibration. Correspondence between the
normalized mean relative velocity of fragments ($\beta^{(N)}_{rel}$) and the
radial collective energy $E_{R}$; full squares (Xe+Sn QF sources from 32 to 50
AMeV incident energies) with error bars are used to
establish the correspondence (dotted line) between $\beta^{(N)}_{rel}$ and
$E_{R}$.}\label{erad1}
\end{figure}

\begin{figure}[!hbt]
\includegraphics*[scale=0.65]{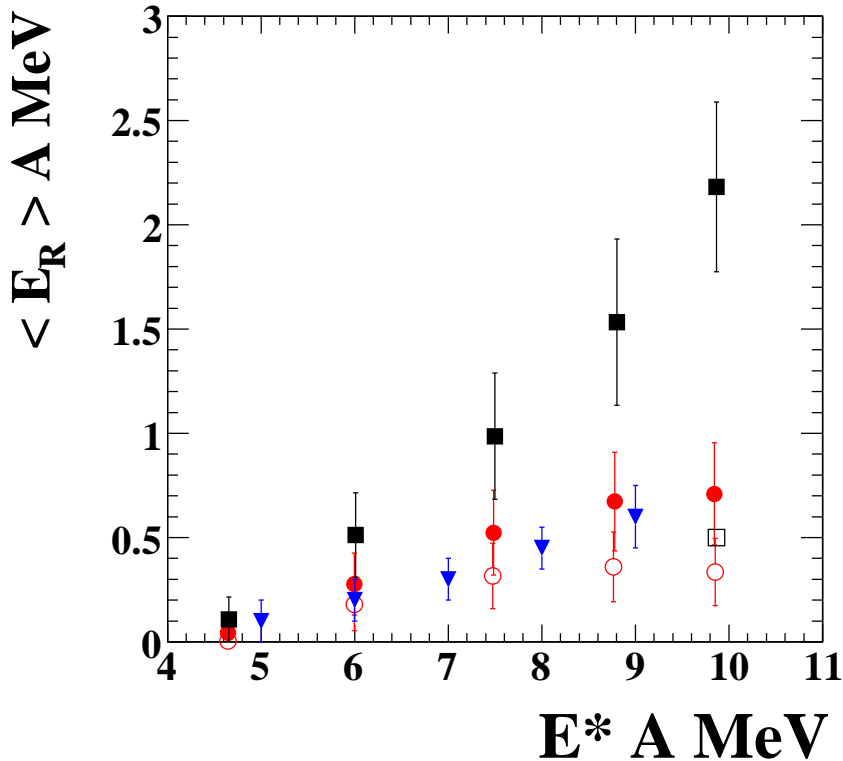}
\caption{Radial collective energy. See fig.~\ref{vr} for symbols for QF and
QP sources. Evolution of the radial collective energy with the
excitation energy per nucleon for different sources; full triangles
correspond to $\pi^{-}+Au$ reactions~\cite{Beau01} and the open square 
to an estimate of
the thermal part of the radial collective energy for Xe+Sn sources produced
at 50 AMeV incident energy (see text).}\label{erad2}
\end{figure}

 In the left part of figure~\ref{vr}, the mean evolution of this
observable with the excitation energy is plotted for the two types of 
sources. For QF sources $\beta_{rel}$ exhibits a linear increase with 
excitation energy. 
For QP sources $\beta_{rel}$ remains almost constant along the $E^{*}$
range. Fragment velocities are the results of the composition of at most 
three components: a thermal kinetic part, determined at freeze-out, 
mainly related to the energy deposit in sources; a Coulomb contribution 
dependent on the source sizes and an eventual radial extra energy. 
The effect of the Coulomb contribution can be removed  by using a 
simple normalization (eq.\ref{eq_vrn}) which takes 
into account, event by event,
the Coulomb influence in the velocity space of the mean fragment charge 
($<Z>$) on the complement of the source charge ($Z_{s}-<Z>$).
\begin{eqnarray}
\label{eq_vrn} \beta^{(N)}_{rel} = \frac{\beta_{rel}}{\sqrt{<Z>(Z_{s}-<Z>)}}
\end{eqnarray}
The mean behaviour of the $\beta^{(N)}_{rel}$ observable so defined is shown 
in the right part of figure~\ref{vr}. At
an excitation energy of about 5 AMeV, $\beta^{(N)}_{rel}$ values
corresponding to QF and QP sources are similar.
Above that excitation energy, the values for QF sources exhibit a strong
linear increase. For QP sources $\beta^{(N)}_{rel}$ slightly increases up to 9
AMeV excitation energy and saturates above. 
That saturation can be attributed to the
compact source selection. Indeed if one performs the same analysis of 
QP's (i.e. with the same limit on VarDyn) from Au+Au collisions at 
100~AMeV, $\beta^{(N)}_{rel}$ increases faster 
and saturates at a slightly higher excitation energy, around 10 AMeV 
(full points in the figure).
That fast divergence between the values of
$\beta^{(N)}_{rel}$ for the two types of sources signals the well known 
onset of radial collective expansion for central collisions. 
Indeed in~\cite{I59-Bor04}, estimates of radial collective energy 
(from 0.5 to 2.2 AMeV) for QF sources produced by Xe+Sn collisions are 
reported for four incident energies: 32, 39, 45 and 50 AMeV.
Those estimates were extracted from comparisons with the statistical model
SMM assuming a self similar expansion energy. The four estimated values of 
the radial collective energy ($E_{R}$) for the QF sources can
be used to calibrate the $\beta^{(N)}_{rel}$ observable. 
The correspondence between $\beta^{(N)}_{rel}$ and $E_{R}$ is 
deduced from a second order polynomial
adjustment (dotted line in  figure~\ref{erad1}).
From this function, we extract firstly a radial energy estimate for QF sources
formed at 25 AMeV incident energy: 0.1$\pm$0.1 AMeV; the error bar being
determined from the grey zone in figure~\ref{erad1}. In
the same way we also deduce two sets (80 an 100 AMeV collisions)
of $E_{R}$ values with their error bars for QP sources 
using the appropriate mean values of $\beta^{(N)}_{rel}$. 
For that calibration we take intervals of $E^*$ centered around the mean 
value, with a width equal to $\pm 3 \sigma$, of the 
corresponding QF source excitation energy distributions.
All the quantitative information concerning 
the evolution of radial energy with excitation energy for both types 
of sources 
is presented in figure \ref{erad2}. We have also added the 
$E_{R}$ values published by the ISIS collaboration~\cite{Beau01}
corresponding to the $\pi^{-}+$Au reactions which
provide sources equivalent to the QP ones in terms of excitation energy
range and size. The observed evolution of $E_{R}$ for such sources is almost
the same as for QP sources. For hadron induced reactions the thermal
pressure is the only origin of radial expansion, which indicates that it is
the same for QP sources. To be fully convincing, an estimate of the 
 part of the radial collective energy due to thermal pressure 
 calculated with the EES model~\cite{Fri90} for an excited nucleus 
 identical to QF sources produced at 50~A~MeV incident energy 
is also reported (open square) in the figure~\cite{BBro96}. 
To conclude on radial collective energy  we have shown that it is 
essentially produced by thermal pressure
in semi-peripheral heavy-ion collisions as it is in hadron induced reactions. 
For QF sources produced in central heavy-ion collisions the contribution 
from the compression-expansion cycle becomes more and more important 
as the incident energy increases.

\section{Discussion and conclusions}
We have compared the properties of QF and QP multifragmenting sources in the
excitation energy range 5-10 AMeV. They present similarities and differences.
Similarities concern:  the division of charge
among particles and fragments, the particle multiplicities (both normalized
to the source size), the average charge of the heaviest 
fragment of the partitions and the fluctuations of the normalized fragment
multiplicities, all the mentioned as a function of excitation 
energy (including radial collective energy).
Differences are first relative to the fluctuations in charge/size of 
the heaviest fragment, that are larger for QP than for QF sources. 
That observation was
very recently mentioned~\cite{I63-NLN07} for normalized
quantities $\sigma(Z_1/Z_s$) and a possible explanation, by comparison with
statistical model (SMM) calculations, was related to different freeze-out
volumes. Note that a detailed study
using the recently developed theory of universal $\Delta$ scaling
laws~\cite{Bot01,I51-Fra05,I61-Pic06} should also be of the prime interest
to make progress
on the understanding of the different fluctuations observed. 
Secondly the asymmetry of the
fragment partitions, $A_{Z} \backslash \{Z_{1}\}$, is also larger for QP
sources; that difference was observed at
5 AMeV excitation energy and above, which corresponds to the onset of
collective energy. Thus, the lower asymmetry for QF sources seems to be
related to the presence of radial collective expansion coming from the
compression-expansion cycle for central collisions.
Finally normalized fragment multiplicities for QF sources are also 
significantly larger above 5 AMeV. Clearly the degree of fragmentation of
the system increases with the radial collective energy and partitions are
also affected.
What is the influence of the radial collective energy on the region
where QF and QT sources break in
the phase diagram (plane freeze-out volume - excitation energy)?

\begin{figure}[!hbt]
\includegraphics*[scale=0.65]{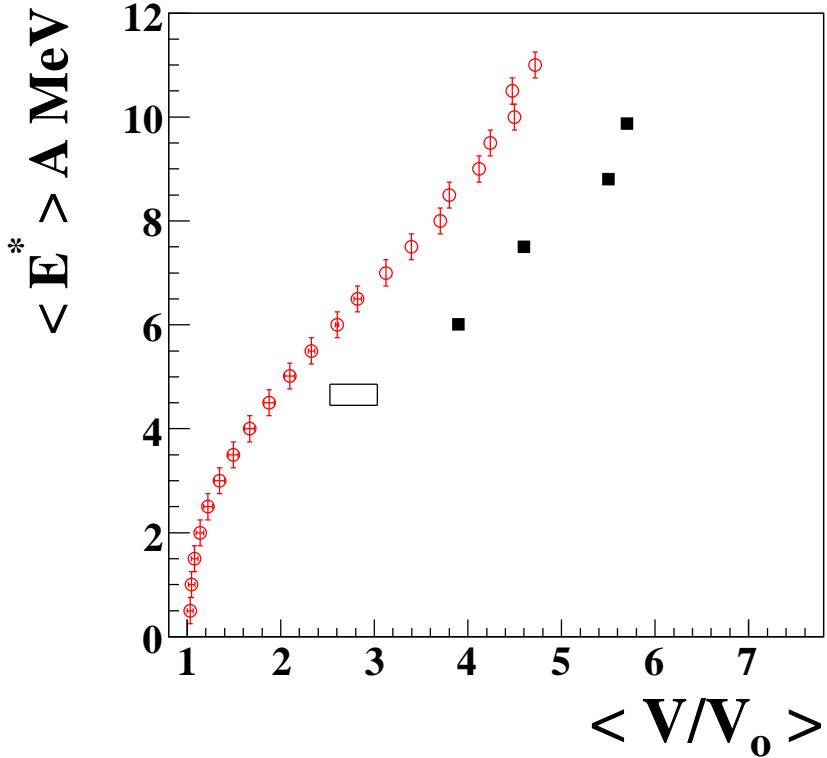}
\caption{Fragmentation position in the excitation energy-freeze-out volume
plane. The four full squares are taken from~\cite{I66-Pia08}. 
The open rectangle gives the estimated position (with error bar) for QF
source at 25~AMeV, and the open circles those for QP sources.}\label{diagph}
\end{figure}

Estimates of freeze-out volumes for QF sources produced 
in Xe+Sn collisions for incident energies between 32 and 50 AMeV
were very recently obtained (details can be found in~\cite{I66-Pia08}).
They evolve from 3.9 to 5.7 $V/V_0$, where $V_0$ would correspond to 
the volume of the source at normal density. 
These results were extracted from simulations using experimental data 
and agree rather well with those deduced from the
statistical model MMM~\cite{Rad97,Radu00}.
In a first attempt to calibrate the freeze-out volumes for other sources,
we use the charge of the heaviest fragment $<Z_1^{(N)}>$ or the 
fragment multiplicity $<M_{frag}^{(N)}>$, normalized to the size of the 
source, as representative of the volume or density at break-up.
From the four points for QF sources and the additional constraint that
$Z_1^{(N)} = M_{frag}$=1 at $V/V_0$=1, we obtain two relations
$V/V_0 = f_1(Z_1^{(N)} )$ and  $V/V_0 = f_2(M_{frag}^{(N)})$, from which we
calculate the volumes for QF sources at 25~AMeV and for QP sources.
The results are plotted in fig.~\ref{diagph}, with error bars coming
from the difference between the two estimates using $f_1$ and $f_2$; note that 
error bars for the QP volumes are small up to 7~AMeV, and can not be 
estimated above, due to the fall of $<M_{frag}^{(N)}>$ at high energy (see
fig.~\ref{zs_zf}).

The volumes of QP sources are smaller than those of QF sources
(about $1.2 \times V/V_0$ for $E^*$=10~AMeV). This supports the observation 
made previously starting from fluctuations of the charge of the 
heaviest fragment in a partition (see figures 13 and 18 in~\cite{I63-NLN07}). 
It is worth noting that a freeze-out volume 
significantly larger than that of a QP source (2.75$\pm0.25 V/V_0$ 
compared to 1.9 $V/V_0$) at the same excitation energy per nucleon 
is found for QF sources at 25~AMeV. This could indicate that as soon as
compression-expansion occurs, larger volumes are involved.

To conclude, we have compared in detail the static properties of hot 
fragmenting
sources produced in central (QF) and semi-peripheral (QP) collisions. From
kinematical properties of fragments, information on the different radial
collective energies involved was deduced using mean fragment relative
velocities, a comparison with hadron-nucleus results and an estimate of the
part of the radial collective energy due to thermal pressure. 
The major results are the following. The weak radial collective
energy observed for QP sources is shown to originate from thermal pressure
only; it reaches about 0.7 AMeV at an excitation energy of
10 AMeV. The larger fragment multiplicities observed for QF sources 
and their more symmetric fragmentation must be related to the extra 
radial collective energy produced by the compression-expansion cycle 
occurring in central collisions.
Such a cycle seems to lead fused systems to break at lower density.

\section*{Appendix}
The values reported in this paper for the excitation energies of the hot
sources sometimes differ from those published by the INDRA collaboration for
the same systems. In this appendix we will examine the influence of the
parameters entering the calorimetric equation on the excitation energy per
nucleon of the sources.

The excitation energy, $E^*$, of a hot source is calculated event 
by event with the relation
\begin{equation}
E^*_s = \sum_{M_{cp}} E_{cp} + \sum_{M_n} E_{n} - Q.
\end{equation}
$M_{cp}$, $E_{cp}$ and $M_{n}$, $E_{n}$ are respectively the multiplicities
and kinetic energies of charged
products and neutrons belonging to the source; $Q$ is the mass difference
between the source and all final products.  
Energies are expressed in the source reference frame. 

We consider Xe+Sn QF sources, for which the
reference frame is the well defined reaction centre of mass,
and discuss only the average values (we know that the procedure
used for calorimetry broadens the distributions).
We have firstly verified  that the definition of a fragment
(Z$\geq$3 or Z$\geq$5) has no influence (less than 0.05 AMeV)
on the excitation energy of the source.

As for all results obtained with the INDRA array, neutrons are not 
detected. Their multiplicity is equal to the difference between the 
mass of the source and the sum of those of the final products. 
The source is assumed to have the same N/Z ratio as the initial 
system (projectile + target). The fragment masses 
are derived in the present paper from the EAL estimate. 
In other papers, the mass of the $\beta$-stability valley
is chosen~\cite{I63-NLN07}. These two mass formulae differ for large 
fragments $Z>20$. However due to rounding up compensations for the 
light fragment masses, it appears that, when using the $\beta$-stability 
mass, the neutron number is 2 units (7\%) larger 
 at 32 AMeV and 5 units (20\%) larger  at 50 AMeV. 
 The Q-value also depends on 
the neutron multiplicity, at the level of a few percents.

The kinetic energy of neutrons is calculated by firstly assuming that the
source has a temperature T given by:
$E^*_s = (A_s/k) T^2$,
which, introduced in the former equation gives:

\begin{equation}
\label{eq_caloQF} E^*_s = \sum_{M_{cp}} E_{cp} + {M_n} fT - Q.
\end{equation}

If the neutrons are all emitted at freeze-out, the factor $f$ is equal to
1.5, whereas it is equal to 1 if all neutrons are evaporated along a long
chain~\cite{Gon89}.
Obviously, the final calculated excitation energy will be higher when $f$ is
larger. The level density parameter plays also a role in the determination
of $E^*$, the value of which slightly increasing with $k$. 

Finally we give in the following table the values of the different terms 
of eq.~\ref{eq_caloQF} with the two sets of hypotheses used in the 
present paper (second and fourth rows)
and in ref.~\cite{I63-NLN07} (third and fifth rows).

\begin{table}[htb]
\begin{center}
\begin{tabular}{|c|c|c|c||c||c|c|c|c|c|}
\hline
E$_{inc}$ & Z$_{frag}^{min}$ & A$_f$ & k & $f$ & $\sum E_{cp}$ & Q & T & M$_n$ & E$^*$ \\
\hline
50 &5 & EAL & 10 & 1.0   & 6.10 & 2.34 & 9.92 & 26.6 & 9.88  \\
50 &3 & $\beta$ & 8 & 1.5& 6.24 & 2.46 & 9.38 & 31.9 & 11.04 \\
\hline
32 & 5 & EAL & 10 & 1.0  & 3.62 & 1.39 & 7.73 & 26 & 6.01 \\
32 & 3 & $\beta$ & 8 & 1.5& 3.72 & 1.40 & 7.27 & 27.9 & 6.64 \\ \hline
\end{tabular}
\caption{Calorimetry results for central Xe+Sn reactions with two 
different sets of hypotheses. The left part lists the hypotheses, 
Z$_{frag}^{min}$ is the minimum fragment charge. 
The right part gives the results. 
All values are in MeV per nucleon, except T (in MeV). The events
considered have a total detected charge equal to at least 80\% of
the system charge.}
\end{center}
\end{table}
If we use all sets of data obtained by varying the fragment mass 
(EAL, $\beta$-stability), the factor $f$ (1.0 and 1.5) and  the 
level density parameter ($k$=8,10), we end-up with values of the 
excitation energies per nucleon which differ at
most by 1.2~MeV at 50 AMeV and 0.8~MeV at 32 AMeV. This helps putting 
systematic error bars on the excitation energy, $\pm$6.5\% at 32 AMeV 
and $\pm$6\%  at 50 AMeV.
These systematic error bars are in agreement with the estimates 
of~\cite{MDA02,H5Vie06}. Note that the relative values of
$E^*/A$ at the different bombarding energies are little modified.


\end{document}